\documentclass[lettersize,journal]{IEEEtran}
\usepackage{amsmath,amsfonts}
\usepackage{array}
\usepackage{textcomp}
\usepackage{stfloats}
\usepackage{url}
\usepackage{verbatim}
\usepackage{graphicx}
\usepackage{cite}
\usepackage{multirow} 
\usepackage{bm}

\usepackage{graphicx} 
\usepackage[linesnumbered,ruled,vlined]{algorithm2e}
\usepackage{amsmath} 
\usepackage{amssymb}

\usepackage{array}

\usepackage{enumerate}
\usepackage{enumitem}
\usepackage{makecell}
\usepackage[labelformat=empty]{subcaption}
\usepackage[belowskip=3pt, aboveskip=3pt]{caption}
\usepackage{booktabs}

\begin{document}

\title{Pre-trained Prompt-driven Semi-supervised Local Community Detection}

\author{Li Ni, Hengkai Xu, Lin Mu, Yiwen Zhang*, and Wenjian Luo, ~\IEEEmembership{Senior Member,~IEEE}
\thanks{
Li Ni is with Key Laboratory of Intelligent Computing \& Signal Processing, Ministry of Education and School of Computer Science and Technology, Anhui University, Hefei, Anhui, 230601, China.
Hengkai Xu, Lin Mu, and Yiwen Zhang are with the School of Computer Science and Technology, Anhui University, Hefei, Anhui, 230601, China.

Wenjian Luo is with the School of Computer Science and Technology, Harbin Institute of Technology, Shenzhen, 518055, China.

Email: nili@ahu.edu.cn, xhk@stu.ahu.edu.cn, mulin@ahu.edu.cn, zhangyiwen@ahu.edu.cn, and luowenjian@hit.edu.cn. (Corresponding author: Yiwen Zhang)}
\thanks{}}

\markboth{}%
{}

\IEEEpubid{}

\maketitle

\begin{abstract}

Semi-supervised local community detection aims to leverage known communities to detect the community containing a given node.
Although existing semi-supervised local community detection studies yield promising results, they suffer from time-consuming issues, highlighting the need for more efficient algorithms. 
Therefore, we apply the “pre-train, prompt” paradigm to semi-supervised local community detection and propose the Pre-trained Prompt-driven Semi-supervised Local community detection method (PPSL).
PPSL consists of three main components: node encoding, sample generation, and prompt-driven fine-tuning. Specifically, the node encoding component employs graph neural networks to learn the representations of nodes and communities. 
Based on representations of nodes and communities, the sample generation component selects known communities that are structurally similar to the local structure of the given node as training samples. 
Finally, the prompt-driven fine-tuning component leverages these training samples as prompts to guide the final community prediction. 
Experimental results on five real-world datasets demonstrate that PPSL outperforms baselines in both community quality and efficiency.
\end{abstract}


\section{Introduction} \label{introduction}
Community structures are commonly observed in real networks \cite{Yang_02, Li_02, Ni_05}.
Identifying community structures in networks has broad applications for various fields \cite{Liu_03, Tang_01, Guan_01, Zhou_01}, such as group recommendation \cite{Wu_03, Cao_01}, fraud detection \cite{Fionda_01, Madi_01}, and event organization \cite{Qiyuan_01}.
In real-world scenarios, network data often contains some prior information, such as pairwise constraints \cite{Eaton_01, Zhang_02, Li_01, Ganji_01, Berahmand_01}, node labels \cite{Liu_01, Dong_01}, and known community structures \cite{Bakshi_01, Zhang_01, Wu_01, Wu_02}.
To improve the quality of community detection using such prior information, various semi-supervised community detection methods have been proposed \cite{Wu_01}.
For example, Bespoke \cite{Bakshi_01} utilized the size and structure of known communities to identify new communities.
SEAL \cite{Zhang_01} employs a generative adversarial network to extract a general community structure from known communities and generate new communities based on the extracted structure.
Wu et al. \cite{Wu_02} adopt “pre-train, prompt” paradigm to
leverages a few known communities to identify  communities.
%

The above methods focus on detecting communities similar to the structure extracted from known communities. 
However, due to differences in community structures, the structure extracted from known communities may differ from the community structure of a given node. As a result, the detected communities may not include the given node. To address this, Ni et al. proposed a semi-supervised local community detection problem, which aims to identify the community of a given node using known communities \cite{Ni_01, Ni_02, Ni_03, Ni_04}. 
They further developed SLSS \cite{Ni_01}, leveraging structural similarity between communities to guide the identification of a given node's community. Additionally, they introduced SLRL \cite{Ni_04} based on reinforcement learning, which selects communities that are similar to the local structure of the given node to distill useful communities.
Chen et al. \cite{Chen_01} proposed CommunityAF, which integrates an autoregressive flow generation model with graph neural networks (GNN) to learn node embedding and generate communities.
Although existing methods yield promising results, they suffer from time-consuming issues. These inefficiencies arise from frequent subgraph similarity calculations in SLSS, redundant training in SLRL, and frequent incremental GNN updates in CommunityAF. Consequently, there is a clear demand for more efficient algorithms.

In response to this need, we apply the “pre-train, prompt” paradigm to the semi-supervised local community detection task and propose the Pre-trained Prompt-driven Semi-supervised Local community detection (PPSL) model.  
PPSL reduces the number of community expansion steps and leverages a small number of training samples as prompts for efficient training and fast fine-tuning.
It consists of three key components: node encoding, sample generation, and prompt-driven fine-tuning.
The node encoding component leverages GNN to learn the representation of nodes and communities, which is then used for both sample generation and prompt-driven fine-tuning. The sample generation component selects communities with structures similar to the local structure of the given node from known communities to form a training set. Finally, the prompt-driven fine-tuning component uses the training set as a prompt to predict the community of the given node.

The main contributions of this paper are summarized as follows:
\begin{enumerate}[topsep=3pt, partopsep=0pt, itemsep=0pt, parsep=0pt, label=(\arabic*), leftmargin=15pt]
    \item Motivated by the strong performance of the “pre-train, prompt” paradigm in community-related tasks, we introduce it to the semi-supervised local community detection. To the best of our knowledge, this is the first work to adopt the “pre-train, prompt” paradigm to semi-supervised local community detection.
    
    \item We propose a semi-supervised local community detection method based on “pre-train, prompt”, called PPSL. 
    It reduces the number of community expansion steps and leverages a small number of training samples as prompts for efficient training and fast fine-tuning.

    \item Extensive experiments on five real-world datasets demonstrate that PPSL outperforms the baseline algorithms in terms of community quality and also achieves faster runtime performance.
\end{enumerate}

The rest of the paper is organized as follows. Section
\ref{related works} summarizes the related works. Section \ref{methodology} presents the problem statement, the PPSL overview, and each component of PPSL. Section \ref{experiment} reports experimental results. Section \ref{conclusion} concludes the paper.

\section{Related works} \label{related works}
This section will introduce related works on semi-supervised community detection, semi-supervised local community detection, and the “pre-train, prompt” paradigm for graph-based tasks.

\subsection{Semi-supervised community detection}
Semi-supervised community detection aims to identify potential community structures in a graph using prior information \cite{Eaton_01, Liu_01, Bakshi_01}. These methods use various types of prior information, including link constraints \cite{Eaton_01, Zhang_02, Li_01, Ganji_01, Berahmand_01}, node labels \cite{Liu_01, Dong_01, Sara_01}, and known community information \cite{Bakshi_01, Zhang_01, Wu_01, Wu_02}. Among the methods that use known community information, a common approach is seed expansion \cite{Bakshi_01, Zhang_01}, which leverages a training set to select seed nodes from the network and then expands the community around the seed nodes. For example, Bakshi et al. \cite{Bakshi_01} introduced Bespoke, which uses community size and structure information from training communities to mine other communities. Zhang et al. \cite{Zhang_01} designed SEAL, which includes a generator that generates communities and a discriminator that forms a generative adversarial network with the generator to validate the accuracy of these communities. Furthermore, Wu et al. \cite{Wu_01} developed CLARE, which includes a locator and a rewriter. The locator identifies communities similar to those in the training set, and the rewriter further refines their structure to improve quality. Additionally, Wu et al. \cite{Wu_02} proposed ProCom, a model for few-shot target community detection. It first pre-trains to capture latent community structures in the network and then leverages a small number of known community samples to identify target communities.
However, unlike the above approaches that fine-tune models using the entire training set, our method fine-tunes the model using only the training samples corresponding to the given node.

\subsection{Semi-supervised local community detection}
Semi-supervised local community detection uses known community information to identify the community to which a given node belongs \cite{Ni_01, Chen_01, Ni_04}. Ni et al. \cite{Ni_01} first introduced the semi-supervised local community detection problem and proposed SLSS. SLSS leverages structural similarity between communities to guide the search for communities of given nodes. Ni et al. \cite{Ni_04} further proposed SLRL, which incorporates reinforcement learning and consists of an extractor and an expander: the extractor captures the local structure of a given node and refines the training set via spectral clustering, while the expander derives the final community. Chen et al. \cite{Chen_01} designed CommunityAF, which employs graph neural networks to learn node embeddings and uses an autoregressive flow generation model to detect communities containing given nodes. However, the above methods do not integrate the “pre-train, prompt” paradigm with the semi-supervised local community detection problem.

\subsection{“Pre-train, prompt” paradigm for graph-based tasks}
The “pre-train, prompt” paradigm was initially widely used in the field of natural language processing \cite{Liu_02, Wei_01}. In this paradigm, instead of modifying the pre-trained model for specific downstream tasks, this approach reformulates tasks through prompt design, aligning them with those addressed during pre-training. In recent years, the “pre-train and prompt” paradigm, known for its expressiveness and flexibility \cite{Sun_02, Wang_01}, has also been widely adopted for graph-based tasks. For example, Sun et al. \cite{Sun_01} proposed the All-in-one multi-task prompting method, which standardizes the format of the graph and language prompts, including prompt tokens, token structures, and insertion patterns, and then leverages prompts to adapt the pre-trained model to downstream tasks. Fang et al. \cite{Fang_01} proposed GPF, a prompt-based general fine-tuning method for pre-training graph neural networks. Instead of manually designing a specific prompt function for each pre-training strategy, GPF adaptively generates prompt graphs for downstream tasks. Although there are works \cite{Wu_02} on global community detection that model the task using the “pre-train, prompt” paradigm, this approach \cite{Wu_02} is not suitable for semi-supervised local community detection.

\section{Methodology} \label{methodology}
This section introduces the problem statement, the overview of the Pre-trained Prompt-driven Semi-supervised Local community detection (PPSL), and the details of the implementation of the PPSL components.
Table \ref{table: Summary of Notations} shows the notations used throughout this paper.

\begin{table}[!t]
  \caption{Important notations.}
  \label{table: Summary of Notations}
  \begin{tabular}{>{\centering\arraybackslash}p{2.4cm} >{\raggedright\arraybackslash}p{5.4cm}}
    \toprule
    Notations & Description\\
     \midrule
     $G=(V,E,A)$ & the graph $G$, the node set $V$, the edge set $E$, and the node feature vector set $A$.\\
     \midrule
     $C_{k}=\left\{c_{1},\cdots ,c_{k}\right\}$ & the known community set $C_{k}$, a community $c_{i},i\in\left[1, k\right]$.\\ 
     \midrule
     $\dot{c}_{v_{0}}$, $c_{v_{0}}$ & the initial community $\dot{c}_{v_{0}}$ for the given node $v_{0}$, the final prediction community $c_{v_{0}}$ for $v_{0}$.\\
     \midrule
     $z(v)$, $z(c)$ & the embedding of node $v$, the embedding of community $c$.\\
     \midrule
     $C_{s}$ & the similar community set.\\ 
     \midrule
     $N_{v}$, $N_{c}$, $\tilde{N} $ & the local structure of node $v$, the local structure of community $c$, and the corrupted local structure.\\ 
     \midrule
     $\text{GNN}_{\theta }(\cdot )$ & the graph neural network with a parameter $\theta$.\\ 
     \midrule
     $\text{PF}_{\phi}(\cdot)$ & the prompt function with a parameter $\phi$.\\ 

  \bottomrule
\end{tabular}
\end{table}

\subsection{Problem statement}

Given a graph $G=(V, E, A)$, a query node $v_{0} \in V$, and a set of known communities $C_{k}=\{c_{1},\cdots, c_{k}\}$, the goal of the semi-supervised local community detection problem is to predict the community $c_{v_{0}}$ of the query node $v_{0}$ using the communities in $C_{k}$.


\subsection{PPSL overview}

\begin{figure}[!t]
  \centering
  \includegraphics[width=\columnwidth]{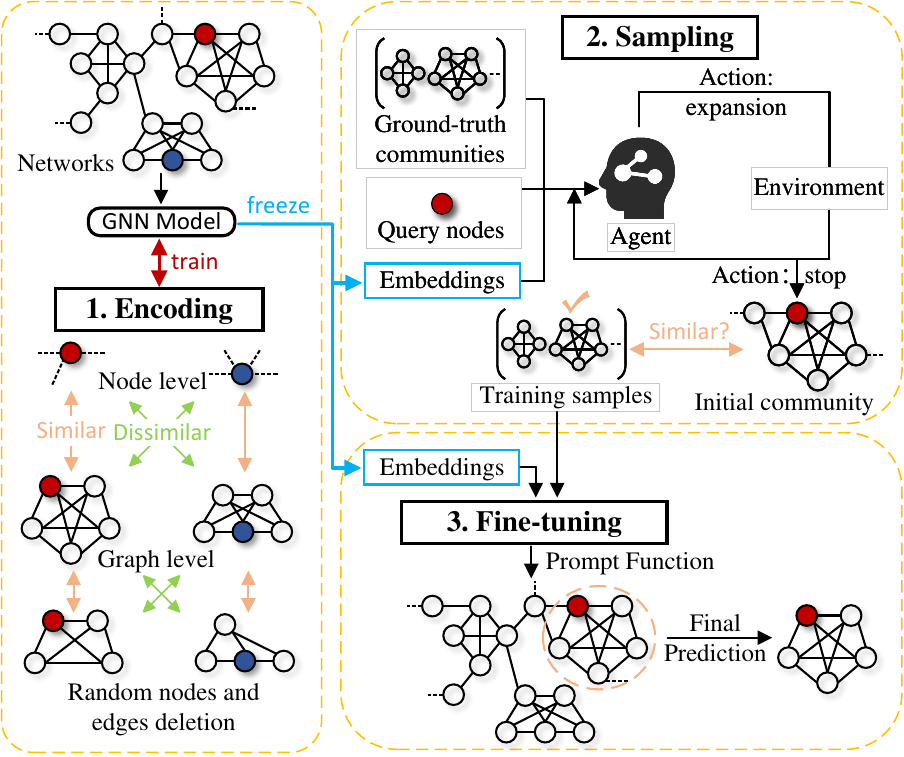}
  \caption{Illustration of PPSL.} 
  \label{figure: PPSL}

\end{figure}

Following the “pre-train, prompt” paradigm, our model consists of three components: node encoding, sample generation, and prompt-driven fine-tuning. The detailed PPSL procedure is summarized in Algorithm \ref{Algo: PPSL Pipeline} and illustrated in Figure \ref{figure: PPSL}.  
In the node encoding component (line~\ref{PPSL: line1}), PPSL utilizes graph neural networks (GNN) and contrastive learning to generate node embeddings that capture the local topological features of nodes and their $k$-ego networks. These embeddings are subsequently leveraged to facilitate downstream community prediction.
Once trained, the GNN encoder outputs node embeddings $z(v) \in R^{d}$ for each node $v \in V$. 

Next, in the sample generation component (line~\ref{PPSL: line2}), we initialize the agent using the node embeddings obtained from the node encoding component, and train it via reinforcement learning based on the known community set $C_{k}$. After training, given a query node $v_0 \in V$, the agent generates its initial community $\dot{c}_{v_0}$. Then, using the GNN encoder from the node encoding component, we obtain the embedding representations of the initial community $\dot{c}_{v_0}$ and each training community $c^i \in C_k$, denoted as $z(\dot{c}_{v_0})$ and $z(c^i)$, respectively. We compute the similarity between $z(\dot{c}_{v_0})$ and each $z(c^i)$ based on the Euclidean distance, and select the $m$ most similar communities from $C_k$ to form a set of similar communities $C_s$, which will serve as the training samples for the subsequent fine-tuning.

\begin{algorithm} [!t]
 \caption{PPSL}
 \label{Algo: PPSL Pipeline} 
\SetKwData{Left}{left}
\SetKwData{This}{this}
\SetKwData{Up}{up} 
\SetKwInOut{Input}{input}
\SetKwInOut{Output}{output}
	\Input{A graph $G=(V, E, A)$, a known community set $C_{k}$, a query node $v_{0}$} 
	\Output{A community $c_{v_{0}}$}
        Conduct \textit{Node encoding} on graph $G$ (Section \ref{node encoding})\; \label{PPSL: line1}
        Perform \textit{Sample generation} for the query node $v_{0}$ (Section \ref{sample generation})\; \label{PPSL: line2}
        Implement \textit{Prompt-driven fine-tuning} (Section \ref{prompt-driven fine-tuning})\; \label{PPSL: line3}
        Generate a final prediction community $c_{v_{0}}$\; \label{PPSL: line4}
        \textbf{return} $c_{v_{0}}$\; \label{PPSL: line5}

 \end{algorithm}

Finally, in the prompt-driven fine-tuning component (line~\ref{PPSL: line3}), the model fine-tunes the prompt function $\text{PF}_{\phi}(\cdot)$ using the training samples, and then performs the final community prediction. First, PPSL initializes $\text{PF}_{\phi}(\cdot)$, implemented as a multilayer perceptron (MLP), using node embeddings obtained from the node encoding component. The prompt function is trained on the node embeddings of the sampled communities $C_s$, with its parameters $\phi$ optimized via a cross-entropy loss. The objective is to ensure that the nodes within the same community exhibit greater similarity in their local structural patterns with respect to their corresponding central nodes. After training, the $k$-ego network of the initial community, denoted as $N_{\dot{c}_{v_0}}$, is put into the prompt function to obtain the final prediction community: $c_{v_0} = \text{PF}_{\phi}(N_{\dot{c}_{v_{0}}})$ (line~\ref{PPSL: line4}).

\subsection{Node encoding}\label{node encoding}

In the node encoding component, which is similar to ProCom \cite{Wu_02}, our model leverages local structure in a graph to train the GNN-based encoder $\text{GNN}_{\theta}(\cdot)$. The objective is to enable the GNN encoder to capture rich topological information of nodes' local structure within a graph, laying the foundation for downstream community prediction tasks. Specifically, the GNN encoder is pre-trained by learning: 1) the similarity between nodes and their respective local structures, and 2) the similarities and differences among local structures themselves. 
By doing so, the GNN encoder learns the local structural patterns of nodes, which indirectly reveal community-level distinctions across the graph.
A detailed process of node encoding is shown in Algorithm \ref{Algo: node encoding}.

To model the relationship between nodes and their local structural context, PPSL aligns each node's representation with that of its $k$-ego network. Specifically, the model randomly samples a batch of nodes $B \subset V$, and for each node $v \in B$, extracts its local structure $N_v$ consisting of $v$ and its $k$-hop neighbors. After GNN encoding, the model obtain the representations $z(v)$ and $z(N_v) = \sum_{r \in N_v} z(r)$. The alignment is optimized using the loss from  \cite{Wu_02}: 
\begin{equation} 
\label{equ: node encoding_01} 
L_{n \cdot s}(\theta )= \sum_{v \in B} -\log \frac{\exp(sim(z(v), z(N_v))/\tau)}{\sum_{u \in B} \exp(sim(z(v), z(N_u))/\tau)} 
\end{equation} 
where “$n \cdot s$” denotes “node and local structure”, $\tau$ is a temperature hyperparameter, and $\theta$ are the GNN parameters. By maximizing the similarity between a node and its own local structure while contrasting it with others, the encoder learns to preserve community-relevant local patterns.

To enhance the encoder’s ability to capture community-relevant patterns, PPSL further discriminates and aligns local structural representations. For each node $v$, its $k$-ego local structure $N_v$ is perturbed by randomly removing nodes or edges, yielding a variant $\tilde{N}_v$. The encoder is then trained to align the embeddings of $N_v$ and $\tilde{N}_v$ using the following loss from \cite{Wu_02}: 
\begin{equation} 
    \label{equ: node encoding_02} 
    L_{s \cdot s}(\theta )= {\sum_{v \in B}}-\log \frac{\exp(sim(z(N_{v}), z(\tilde{N}_{v}))/\tau)}{ {\textstyle \sum_{u \in B}\exp(sim(z(N_{v}), z(\tilde{N}_{u}))/\tau)} }  
\end{equation} 
where “$ s\cdot s$” refers to “local structure and local structure”. This contrastive objective helps the model learn more discriminative representations of local structures.

\begin{algorithm} [!t]
 \caption{Node encoding}
 \label{Algo: node encoding} 
\SetKwData{Left}{left}
\SetKwData{This}{this}
\SetKwData{Up}{up} 
\SetKwInOut{Input}{input}
\SetKwInOut{Output}{output}
	\Input{A graph $G=(V, E, A)$} 
	\Output{The encoder $\text{GNN}_{\theta}(\cdot)$}
    Initialize $\text{GNN}_{\theta}(\cdot)$\;
    \ForEach{epoch}{
        Randomly select a batch of nodes $B \subset V$\;
        \ForEach{v $\in$ B}{
            Extract $v$'s local structure $N_{v}=(E_{v}, A_{v})$\;
            Encode $N_{v}$ as $Z_{v} = \text{GNN}_{\theta}(E_{v}, A_{v})$ to obtain node and local structure representations $z(v)$ and $z(N_{v})$\;
            Compute corrupted local structure representation $z(\tilde{N}_{v})$\;
        }
        Update $\theta$ by applying gradient descent to minimize loss based on Equations (\ref{equ: node encoding_01}) and (\ref{equ: node encoding_02})\;
    }
    \textbf{return} $\text{GNN}_{\theta}(\cdot)$\;
 \end{algorithm}

PPSL finalize the node encoding by extracting both node-level and local structure-level features that reflect potential community structures. We input the graph $G=(V, E, A)$ into the trained encoder to obtain node embeddings $Z = \text{GNN}_{\theta}(E, A)$, where $z(v) \in Z$ denotes the initial embedding of node $v$. Community embeddings are computed via sum-pooling: $z(c)={\textstyle \sum_{v \in c}} z(v)$. These embeddings are then used for subsequent sample generation and prompt-driven fine-tuning.

\subsection{Sample generation}\label{sample generation}

\begin{algorithm} [!t]
 \caption{Sample generation}
 \label{Algo: sample generation} 
\SetKwData{Left}{left}
\SetKwData{This}{this}
\SetKwData{Up}{up} 
\SetKwInOut{Input}{input}
\SetKwInOut{Output}{output}
	\Input{A graph $G=(V, E, A)$, a training set $C_{k}$, a query node $v_{0}$, the encoder $\text{GNN}_{\theta}(\cdot)$}
	\Output{Training samples $C_{s}$}
    Initialize agent with all nodes' representation $Z = \text{GNN}_{\theta}(E, A)$\;
    \ForEach{epoch}{
        \ForEach{$c^{i} \in C_{k}$}{
            Get the trajectory $\tau = (S_{0}, a_{1}, r_{1}, \cdots, S_{t})$\;
            Compute the cumulative reward $G_{t}$ for each trajectory based on Equation (\ref{equ: initial_01})\;
        }
        Update the agent's parameter $Q$ based on Equation (\ref{equ: initial_02})\;
        Use Teacher-forcing with Maximum Likelihood Estimation for additional supervision\;
    }
    Generate the initial community $\dot{c}_{v_{0}}$ using the agent\;
    Encode the initial community $z(\dot{c}_{v_{0}})$\; 
    Encode each training sample as $z(c^{i}), c^{i} \in C_{k}$\;
    Use Euclidean distance in the embedding space between each $z(c^{i})$ and $z(\dot{c}_{v_{0}})$ as a similarity measure, retrieve the $m$ most similar training samples $C_{s}$\; 
    \textbf{return} $C_{s}$\;
 \end{algorithm}
 
In the sample generation component, the process consists of two steps. Step one: Based on reinforcement learning, PPSL uses known communities $C_{k}=\{c_{1},\cdots,c_{k}\}$ as the training set to train the agent. The trained agent then generates an initial community $\dot{c}_{v_{0}}$ for a given node $v_{0}$. Step two: our model computes the embedding similarity between the initial community and the known communities using Euclidean distance, and selects the top-$m$ most similar ones as samples.

\subsubsection{Step one}
The agent is designed to iteratively expand a community by selecting appropriate nodes from the neighbors of the current community. Below, we introduce the expansion process and key reinforcement learning terminologies: state, action space, and node representation. Initially, the community for a given node $v_{0}$ is $c_{0}=\{v_{0}\}$. At the beginning of $t$-th expansion, the community is $c_{t-1}=\{v_{0},u_{1},\cdots, u_{t-1}\}$, with the current state denoted as $S_{t-1} = c_{t-1} \cup N(c_{t-1})$, and the action space as $N(c_{t-1})$, where $N(c_{t-1})$ represents the neighbors of $c_{t-1}$. The agent selects a node $u_{t}$ from $N(c_{t-1})$ based on the policy network $\pi(a_{t} | S_{t-1})$, expanding the community to $c_{t}=\{ v_{0},u_{1},\cdots, u_{t}\}$.

To extract features from a given node and its current community, the node representation at step $t$ is initialized based on whether the node is the query node or belongs to the current community \cite{Zhang_01, Ni_04}:
\begin{equation}
    x_{t}^{(0)}(u)=g(u) \cdot q_{1}+I(u \in c_{t-1}) \cdot q_{2}
\end{equation}
\begin{equation}
    g(u) =\left\{\begin{matrix}
        max(z(u)) & u=v_{0} \\
        0 & u\ne v_{0}
        \end{matrix}\right.
\end{equation}
where $x_{t}^{(0)}(u)$ represents the embedding of node $u$ at first iteration in the agent. The function $g(u)$ returns the maximum value of the initial embedding $z(u)$ from the node encoding component if $u$ is the query node; otherwise, it returns 0. The indicator function $I$ equals 1 if all its inputs are true, and 0 otherwise. Parameters $q_{1}, q_{2} \in R^{d}$ control the contributions of different node types. Subsequently, the node representation is processed by iGPN \cite{Ni_04}, resulting in the stacked node representation $X_{t} = \text{iGPN}(X_{t}^{(0)})$.

The model initializes the reinforcement learning agent with node embeddings obtained from the node encoding component to accelerate policy convergence. By providing the agent with informative representations from the beginning, we significantly reduce ineffective exploration and enable faster identification of community-related patterns. This embedding-based initialization focuses the agent's learning process on meaningful structures in the graph, leading to quicker and more efficient policy optimization.

During the reinforcement learning phase, PPSL adopts the policy gradient approach to optimize cumulative rewards. For each training community, a trajectory $\tau = (S_{0}, a_{1}, r_{1}, \cdots, S_{t})$ is generated, where $S$, $a$, and $r$ represents the state, action, and reward. Specifically, each action is assigned a reward, which is defined as the change in F-score between $c_{t}$ and $c_{t-1}$ \cite{Wu_01}:
\begin{equation}
    r_{t} = f(c_{t}, c_{true})-f(c_{t-1}, c_{true})
\end{equation}
where $c_{t}$ and $c_{t-1}$ is the community at step $t$ and $t-1$, and $c_{true}$ represents the corresponding ground-truth community.
The cumulative reward is computed as follows \cite{Ni_04}:
\begin{equation} 
\label{equ: initial_01} 
G_{t} = \sum_{k = 0}^{T-t} \gamma ^{k}\cdot r_{t + k}
\end{equation}
where $\gamma$ is the discount factor. Using the state-action-reward tuples $(S, a, G)$, the policy parameters $Q$ are updated as follows \cite{Wu_01}:
\begin{equation} 
\label{equ: initial_02} 
    Q = Q + lr \cdot \sum_{t = 1}^{T}\nabla \log (\pi(a_{t}|S_{t-1})) \cdot G_{t} 
\end{equation}
where $lr$ denotes the learning rate. After training and optimization, the initial community $\dot{c}_{v_{0}}$ for the query node $v_{o}$ can be obtained.
Except for the initial embedding, the remaining components, including the reward design, forward propagation, policy network, and optimization process, are implemented following the method described in \cite{Ni_04}.

\subsubsection{Step two}
If the initially generated community is directly used as the final prediction, the performance is often suboptimal. This is because the model is trained over the entire training set, where some community structures may significantly differ from that of the query node.

Training samples that are structurally similar to the initial community can better guide the model toward accurate community prediction \cite{Ni_01, Ni_04}.  
To address this issue, we construct a customized training set for each query node by selecting structurally relevant samples.
Specifically, given the initial generated community $\dot{c}_{v_0}$ of the query node and the training communities $c^{i} \in C_k, i \in \{1, 2, \dots\}$, we first obtain their embeddings using the GNN encoder from the node encoding component, denoted as $z(\dot{c}_{v_0})$ and $z(c^i)$, respectively.  
The similarity between communities is then measured by the Euclidean distance in the latent space, calculated as:
\begin{equation}
    \text{Sim}(\dot{c}_{v_0}, c^i) = -\left\| z(\dot{c}_{v_0}) - z(c^i) \right\|_2
\end{equation}
where a higher value of $\text{Sim}(\dot{c}_{v_0}, c^i)$ indicates greater similarity.

Based on the computed similarities, we select the top-$m$ most similar communities from $C_k$ to form a set of similar communities, denoted as $C_s$. This tailored set $C_s$ serves as the support samples for subsequent prompt-based learning.  
By selecting communities that are structurally similar to the query node, the model can learn from more relevant samples, which helps it make more accurate predictions.
The detailed procedure for sample generation is summarized in Algorithm~\ref{Algo: sample generation}.

\subsection{Prompt-driven fine-tuning}\label{prompt-driven fine-tuning}

\begin{algorithm} [!t]
 \caption{Prompt-driven fine-tuning}
 \label{Algo: prompt-driven fine-tuning} 
\SetKwData{Left}{left}
\SetKwData{This}{this}
\SetKwData{Up}{up} 
\SetKwInOut{Input}{input}
\SetKwInOut{Output}{output}
    \Input{A graph $G=(V, E, A)$, training samples $C_{s}$, the encoder $\text{GNN}_{\theta}(\cdot)$} 
    \Output{A Prompt Function $\text{PF}_{\phi}(\cdot)$}
    Obtain all nodes' representation as $Z = \text{GNN}_{\theta}(E, A)$\;
    Initialize Prompt Function $\text{PF}_{\phi}(\cdot)$\;
    \ForEach{epoch}{
        Pick a node $v$ from a randomly sampled prompt community $C_{s}$\;
        Extract $v$'s local structure $N_{v}$\;
        Get node embeddings $\{z(u)\}_{u \in N_{v}}$ from $Z$\;
        Update $\phi$ by applying gradient descent to minimize loss based on Equation (\ref{equ: prompt_01})\;
    }
    \textbf{return} $\text{PF}_{\phi}(\cdot)$\;
 \end{algorithm}

To enhance the efficiency of identifying the community of a query node, PPSL incorporates a prompt-driven fine-tuning mechanism \cite{Wu_02}. This mechanism leverages a small number of similar communities as prompts to guide the extraction of the final community from the given node’s local structure.

Specifically, for a given query node $v_0$, PPSL first extracts a $k$-ego local structure $N_{\dot{c}_{v_0}}$ of initial community $\dot{c}_{v_0}$. The model then applies a prompt function $\text{PF}_{\phi}(\cdot)$ to identify the final prediction community $c_{v_0}$ from this local structure. The prompt function is implemented as a learnable multilayer perceptron (MLP) that estimates the similarity between $v_{0}$ and each node $u$ in $N_{\dot{c}_{v_{0}}}$. If this similarity exceeds a threshold $\alpha$, $u$ is added to $c_{v_{0}}$ \cite{Wu_02}:
\begin{equation}
    c_{v_{0}} = \{u \in N_{\dot{c}_{v_{0}}} \mid \sigma(\text{MLP}_{\phi}(z(u) || z(v_{0}))) \ge \alpha \}
\end{equation}
where $z(u)$ and $z(v_{0})$ denote the embeddings of node $u$ and $v_{0}$, respectively, and $\sigma(\cdot)$ is the sigmoid function.

To train the prompt function $\text{PF}_{\phi}(\cdot)$, PPSL utilizes a set of similar communities $C_{s}$ as supervision. For each community $c_{i} \in C_{s}$, a central node $w \in c_{i}$ is randomly selected, and its local structure $N_{w}$ is extracted. Each node $u \in N_{w}, u \ne w$, is labeled based on whether it belongs to $c_{i}$. This binary label is used as the supervisory signal for training the MLP. The objective function is defined as \cite{Wu_02}:
\begin{equation}
\label{equ: prompt_01}
    \begin{split}
        L_{PF}(\phi) &= \sum_{i = 1}^{m} \sum_{w \in c_{i}, u \in N_{w}} I(u, c_{i}) \log \sigma (\text{MLP}(z(u)||z(w))) \\
        &\quad + (1 - I(u, c_{i})) \log (1 - \sigma (\text{MLP}(z(u)||z(w))))
    \end{split}
\end{equation}
where $I(u, c_{i})$ is the indicator function that returns 1 if $u \in c_{i}$ and 0 otherwise.

Notably, since every node in a training community $c_{i}$ can serve as a central node, even a small training set can generate a large number of supervisory signals. This facilitates effective optimization of $\text{PF}_{\phi}(\cdot)$. Once trained, the model applies $\text{PF}_{\phi}(\cdot)$ to the local structure $N_{\dot{c}_{v_{0}}}$ to obtain the predicted community $c_{v_{0}}$. A detailed process of prompt-driven fine-tuning is shown in Algorithm \ref{Algo: prompt-driven fine-tuning}.

In the community generation phase after model training, unlike previous semi-supervised local community detection methods \cite{Ni_01, Ni_04}, PPSL does not rely on iterative community expansion. Instead, it directly identifies the final community using prompt-guided inference, thereby improving prediction efficiency.

\section{Experimrnt} \label{experiment}

In this section, we present the experimental settings, including datasets, evaluation metrics, and comparison algorithms. Additionally, we conduct extensive experiments to evaluate the model's performance on five real-world datasets.

\subsection{Settings}

\begin{table*}[!t]
  \caption{Statistics of datasets. The first two columns are the number of nodes and edges. “$\#c$” and “$|\overline{c}|$” denote the number and the average size of communities in the corresponding dataset.}
  \renewcommand\arraystretch{1.1}
  \label{table: Description of Datasets}
  \begin{center}
  \small
  \setlength{\tabcolsep}{7mm}{
  \begin{tabular}{ccccc}
      \toprule
      Datasets & Nodes & Edges & $\#c$ & $|\overline{c}|$  \\
      \midrule
      \itshape{Amazon} & 13,178 & 33,767 & 4,517 & 9.31 \\
     \itshape{DBLP} & 114,095 & 466,761 & 4,559 & 8.4 \\
     \itshape{Twitter} & 87,760 & 1,293,985 & 2,838 & 10.88 \\
     \itshape{Youtube} & 216,544 & 1,393,206 & 2,865 & 7.67 \\
     \itshape{LiveJournal} & 316,606 & 4,945,140 & 4,510 & 17.65 \\
     \bottomrule
  \end{tabular}}
  \end{center}
\end{table*}

\begin{table}[!t]
    \centering
    \small
    \caption{Hyper-parameter settings for encoding, sampling, and fine-tuning component.}
    \label{table: hyper-parameter}
    \resizebox{\columnwidth}{!}{%
    \begin{tabular}{c|c|c}
        \toprule
        Component & Hyper-parameter & Value \\
        \midrule
        \multirow{9}{*}{Encoding} 
        & Batch size & 256 \\
        & Number of epochs & 30 \\
        & Learning rate & 1e-3 \\
        & Implementation of $\text{GNN}_\Theta(\cdot)$ & 2 layers GCN \\
        & $k$-ego subgraph & 2 \\
        & Embedding dimension & 128 \\
        & Temperature $\tau$ & 0.1 \\
        & Ratio $\rho$ for corruption & 0.85 \\
        & Loss weight $\lambda$ & 1 \\
        \midrule
        \multirow{8}{*}{Sampling} 
        & Batch size & 32 \\
        & Number of epochs & 100 \\
        & Embedding dimention & 64 \\
        & MLP layers & 3 \\
        & iGPNs layers & 3 \\
        & Learning rate & 1e-2 \\
        & Discount factor $\gamma$ & 1 \\
        \midrule
        \multirow{6}{*}{Fine-tuning} 
        & Implementation of $\text{PF}_\Phi(\cdot)$ & 2 layers MLP \\
        & Number of epochs & 30 \\
        & Learning rate & 1e-3 \\
        & $k$-ego subgraph & 3 \\
        & Number of prompts $m$ & 20 \\
        & Threshold value $\alpha$ & 0.2 \\
        \bottomrule
    \end{tabular}
    }
\end{table}

\textbf{Datasets}. We use five real-world datasets from SNAP\footnote{https://snap.stanford.edu/data/}, including Amazon \cite{Yang_01}, DBLP \cite{Backstrom_01}, Twitter \cite{McAuley_01}, Youtube \cite{Mislove_01}, and LiveJournal \cite{Backstrom_01}. The details of these datasets are summarized in Table \ref{table: Description of Datasets}. Since Amazon, DBLP, Youtube, and LiveJournal do not provide node attributes, the initial representations of a node $v$ is defined following \cite{Zhang_01} as: $z(v) = \big[ \deg(v), \max(D_N(v)), \min(D_N(v)), \operatorname{mean}(D_N(v)),\\ \operatorname{std}(D_N(v)) \big]$, where $\deg(v)$ denotes the degree of node $v$ and $D_N(v)$ is the set of degrees of $v$'s neighboring nodes. For each dataset, we select query nodes from the list at regular intervals following the sampling method in \cite{Ni_04}. Specifically, we sample 1000 nodes from Amazon and DBLP, and 500 nodes from Twitter, YouTube, and LiveJournal. Across all datasets, we set $m$ as 20. The hyperparameter settings of the model are listed in Table \ref{table: hyper-parameter}. Our model is implemented using PyTorch\footnote{https://pytorch.org/} and runs on an NVIDIA RTX 4090D-24G GPU. 

\textbf{Evaluation metrics}. To evaluate the performance of the algorithms, we use Precision \cite{Fisher_01}, Recall \cite{Fisher_01}, F-score \cite{Powers_01}, and Jaccard \cite{Paul_01} to measure the difference between the detected and ground-truth communities. A higher metric value indicates a more accurate community detection result. For the semi-supervised community detection algorithms, we extract the communities containing the query nodes from their detected results for comparison.

\textbf{Comparison algorithms.}

(1) An Unsupervised local community detection algorithm includes the M method.
\begin{itemize}[topsep=3pt, partopsep=0pt, itemsep=0pt, parsep=0pt, leftmargin=20pt]
    \item M \cite{Luo_01}: This method is a local modularity-based approach for identifying community structures in large-scale networks.
\end{itemize}

\begin{table*}[!t]
  \centering
  \caption{Results of M, SLSS, ComAF, SLRL, and PPSL. The best results are highlighted in \textbf{bold}, and the second-best results are underlined.}
  \label{table: local results}
  \small
  \begin{tabular}{
>{\centering\arraybackslash}p{0.1\textwidth} 
>{\centering\arraybackslash}p{0.1\textwidth} 
>{\centering\arraybackslash}p{0.1\textwidth} 
>{\centering\arraybackslash}p{0.1\textwidth} 
>{\centering\arraybackslash}p{0.1\textwidth} 
>{\centering\arraybackslash}p{0.1\textwidth} 
>{\centering\arraybackslash}p{0.1\textwidth}}
    \toprule
    Datasets & Metrics & M & SLSS & CommunityAF & SLRL & PPSL \\
    \midrule
    \multirow{4}{*}{\textit{Amazon}} & Precision & 0.8062 & 0.8464 & 0.8295 & \underline{0.8546} & \textbf{0.8683} \\
    \multirow{4}{*}{\textit{}} & Recall & 0.7922 & 0.8683 & 0.9046 & \textbf{0.9354} & \underline{0.9341}\\
    \multirow{4}{*}{\textit{}} & F-score & 0.7596 & 0.8284 & 0.8446 & \underline{0.8782} & \textbf{0.8827}\\
    \multirow{4}{*}{\textit{}} & Jaccard & 0.6572 & 0.7273 & 0.7504 & \underline{0.7983} & \textbf{0.8151}\\
    
    \midrule
    \multirow{4}{*}{\textit{DBLP}} & Precision & 0.5995 & 0.6176 & 0.5153 & \underline{0.6726} & \textbf{0.6833} \\
    \multirow{4}{*}{\textit{}} & Recall & 0.5963 & 0.7204 & \textbf{0.8016} & 0.7264 & \underline{0.7401}\\
    \multirow{4}{*}{\textit{}} & F-score & 0.5545 & 0.6266 & 0.6063 & \underline{0.6622} & \textbf{0.6684}\\
    \multirow{4}{*}{\textit{}} & Jaccard & 0.4412 & 0.5074 & 0.4592 & \underline{0.5394} & \textbf{0.5492}\\
   
    \midrule
    \multirow{4}{*}{\textit{Twitter}} & Precision & 0.2626 & 0.3178 & \textbf{0.4276} & \underline{0.3573} & 0.3395 \\
    \multirow{4}{*}{\textit{}} & Recall & 0.7474 & \textbf{0.5755} & 0.3438 & 0.5445 & \underline{0.5984}\\
    \multirow{4}{*}{\textit{}} & F-score & 0.3084 & 0.3775 & 0.2818 & \underline{0.3785} & \textbf{0.3943}\\
    \multirow{4}{*}{\textit{}} & Jaccard & 0.2055 & \underline{0.2543} & 0.1792 & 0.2516 & \textbf{0.2678}\\
    
    \midrule
    \multirow{4}{*}{\textit{Youtube}} & Precision & 0.2833 & 0.2471 & 0.3520 & \textbf{0.5394} & \underline{0.4495} \\
    \multirow{4}{*}{\textit{}} & Recall & 0.3046 & \underline{0.3705} & 0.2953 & 0.2724 & \textbf{0.3802}\\
    \multirow{4}{*}{\textit{}} & F-score & 0.2443 & 0.2412 & 0.2774 & \underline{0.2925} & \textbf{0.3177}\\
    \multirow{4}{*}{\textit{}} & Jaccard & 0.1603 & 0.1544 & 0.1807 & \underline{0.1923} & \textbf{0.2142}\\

    \midrule
    \multirow{4}{*}{\textit{LiveJournal}} & Precision & 0.5693 & 0.5814 & \textbf{0.6657} & \underline{0.6654} & 0.5842 \\
    \multirow{4}{*}{\textit{}} & Recall & 0.7268 & \underline{0.7976} & 0.6384 & 0.7025 & \textbf{0.8004}\\
    \multirow{4}{*}{\textit{}} & F-score & 0.6092 & \textbf{0.6431} & 0.6002 & 0.5802 & \underline{0.6174}\\
    \multirow{4}{*}{\textit{}} & Jaccard & \underline{0.5166} & \textbf{0.5314} & 0.4853 & 0.4682 & 0.4974\\
    
  \bottomrule
\end{tabular}
\end{table*}

(2) Semi-supervised local community detection algorithms include SLSS, CommunityAF, and SLRL.
\begin{itemize}[topsep=3pt, partopsep=0pt, itemsep=0pt, parsep=0pt, leftmargin=20pt]
    \item SLSS \cite{Ni_01}: This method is the first semi-supervised local community detection algorithm, which proposes a graph kernel-based community structure similarity measurement.
    \item CommunityAF \cite{Chen_01}: This method is a semi-supervised local community detection model consisting of three components: a GNN-based feature extraction module, an autoregressive flow-based community generation module, and a scoring module.
    \item SLRL \cite{Ni_04}: This method incorporates a reinforcement learning-based community extractor and expander.
\end{itemize}

(3) Semi-supervised community detection algorithms include Bespoke, SEAL, CLARE, and ProCom.
\begin{itemize}[topsep=3pt, partopsep=0pt, itemsep=0pt, parsep=0pt, leftmargin=20pt]
    \item Bespoke \cite{Bakshi_01}: This method is a semi-supervised community detection method that leverages community structure and size information.
    \item SEAL \cite{Zhang_01}: This method is a generative adversarial network-based method designed to learn heuristic rules for identifying communities.
    \item CLARE \cite{Wu_01}: This method is a subgraph-based reasoning model comprising a locator and a rewriter for community detection.
    \item ProCom \cite{Wu_02}: This method searches for target communities using a pre-trained prompt model.
\end{itemize}

The number of communities generated by Bespoke, SEAL, CLARE, and ProCom is set to 5000. Similar to ProCom, our algorithm terminates once the generated community size exceeds the maximum size of any known community. For each dataset, we prepare two sets of known communities as training sets, with each set containing 100 known communities. All semi-supervised methods are tested three times with different random seeds, and the final experimental results are obtained by averaging the evaluation metrics across both training sets.

\subsection{Results}

\subsubsection{Comparison with semi-supervised local community detection algorithms}

We compared the results of M, SLSS, CommunityAF, SLRL, and PPSL. Table \ref{table: local results} presents the experimental results for five datasets.

Experimental results indicate that PPSL outperforms M, SLSS, CommunityAF, and SLRL in terms of overall performance. On the Amazon, DBLP, Twitter, and YouTube datasets, PPSL achieves higher F-score and Jaccard values compared to the other methods, except the LiveJournal dataset, where its performance is slightly inferior. In general, semi-supervised local community detection algorithms (SLSS, CommunityAF, SLRL, and PPSL) tend to surpass the unsupervised method M, as they leverage known community information as prior knowledge, thereby guiding the model to identify community structures more accurately. However, the M method exhibits slightly better performance on the LiveJournal dataset because the community structure in this dataset aligns more closely with the modeling assumptions underlying the M method.

On most datasets other than LiveJournal, PPSL outperforms SLRL. This is primarily due to PPSL's ability to learn structural characteristics of local structures through the node encoding component, which enhances the model's generalization capacity. Additionally, PPSL incorporates a prompt mechanism to generate query-specific community predictions, thereby improving prediction accuracy. In contrast, SLRL performs relatively poorly on the LiveJournal dataset due to its inability to learn effective features with limited labeled data, whereas PPSL maintains strong predictive performance even with few samples. PPSL outperforms CommunityAF because it utilizes query-specific training data during the community prediction process, which effectively reduces interference from unrelated community structures. In comparison, CommunityAF employs a uniform training set for all nodes and lacks the capacity for personalized modeling.

\subsubsection{Comparison with semi-supervised community detection algorithms}

\begin{table*}[!t]
 \centering
 \caption{Results of Bespoke, SEAL, CLARE, ProCom, and PPSL.}
  \label{table: global results}
  \small
  \setlength{\tabcolsep}{3.8mm}{
    \begin{tabular}{cccccccccc}
    \toprule
    \multirow{2}{*}{Datasets} & \multirow{2}{*}{Metrics} & \multicolumn{2}{c}{Bespoke-PPSL} & \multicolumn{2}{c}{SEAL-PPSL} & \multicolumn{2}{c}{CLARE-PPSL} & \multicolumn{2}{c}{ProCom-PPSL}\\
    \cmidrule(lr){3-4} \cmidrule(lr){5-6} \cmidrule(lr){7-8} \cmidrule(lr){9-10}
     & & Bespoke & PPSL & SEAL & PPSL & CLARE & PPSL & ProCom & PPSL\\
    \midrule
    \multirow{5}{*}{\textit{Amazon} }
    & Precision & 0.8167 & \textbf{0.8537} & 0.8385 & \textbf{0.8642} & 0.8385 & \textbf{0.8596} & 0.8299 & \textbf{0.8612}\\
    & Recall & 0.8214 & \textbf{0.9256} & 0.8837  & \textbf{0.9324} & 0.8013 & \textbf{0.9364} & 0.8133 & \textbf{0.9326}\\
    & F-score & 0.8193 & \textbf{0.8905} & 0.8394 & \textbf{0.8975} & 0.7953 & \textbf{0.8965} & 0.8216 & \textbf{0.8952} \\
    & Jaccard & 0.6933 & \textbf{0.8015} & 0.7532 & \textbf{0.8131} & 0.6876 & \textbf{0.8114} & 0.6965 & \textbf{0.8106} \\
    \midrule
    \multirow{5}{*}{\textit{DBLP} }
    & Precision & 0.5833 & \textbf{0.6742} & 0.6403 & \textbf{0.6925} & 0.5968 & \textbf{0.6425} & \textbf{0.6587} & 0.6542\\
    & Recall & 0.6311 & \textbf{0.7512} & 0.6753  & \textbf{0.7484} & 0.6425 & \textbf{0.7436} & 0.6623 & \textbf{0.7451}\\
    & F-score & 0.6062 & \textbf{0.7102} & 0.6256 & \textbf{0.7227} & 0.5965 & \textbf{0.6893} & 0.6607 & \textbf{0.6979}\\
    & Jaccard & 0.4355 & \textbf{0.5516} & 0.4975 & \textbf{0.5654} & 0.4714 & \textbf{0.5253} & 0.4932 & \textbf{0.5341}\\
    \midrule
    \multirow{5}{*}{\textit{Twitter} }
    & Precision & 0.3102 & \textbf{0.3621} & 0.3304 & \textbf{0.3412} & 0.3126 & \textbf{0.3712} & 0.3373 & \textbf{0.3744}\\
    & Recall & 0.2994 & \textbf{0.5883} & 0.2973  & \textbf{0.5722} & 0.3262 & \textbf{0.5926} & 0.3087 & \textbf{0.5905}\\
    & F-score & 0.2623 & \textbf{0.3992} & 0.2701 & \textbf{0.3974} & 0.2856 & \textbf{0.4064} & 0.2933 & \textbf{0.4077}\\
    & Jaccard & 0.1598 & \textbf{0.2836} & 0.1665 & \textbf{0.2723} & 0.1794 & \textbf{0.2964} & 0.1817 & \textbf{0.2984}\\
    \midrule
    \multirow{5}{*}{\textit{Youtube} }
    & Precision & 0.3923 & \textbf{0.4512} & \textbf{0.4463} & 0.4438 & 0.3957 & \textbf{0.4584} & 0.4107 & \textbf{0.4535}\\
    & Recall & 0.3535 & \textbf{0.3894} & 0.3596  & \textbf{0.3947} & 0.3525 & \textbf{0.3814} & 0.3587 & \textbf{0.3927}\\
    & F-score & 0.3174 & \textbf{0.3825} & 0.3568 & \textbf{0.3876} & 0.3225 & \textbf{0.3864} & 0.3277 & \textbf{0.3858}\\
    & Jaccard & 0.2065 & \textbf{0.2667} & 0.2494 & \textbf{0.2645} & 0.2174 & \textbf{0.2635} & 0.2216 & \textbf{0.2653}\\
    \midrule
    \multirow{5}{*}{\textit{LiveJournal} }
    & Precision & 0.6315 & \textbf{0.6533} & \textbf{0.7592} & 0.6512 & \textbf{0.6914} & 0.6423 & \textbf{0.6732} & 0.6573\\
    & Recall & 0.6527 & \textbf{0.7856} & 0.6815  & \textbf{0.7844} & 0.6585 & \textbf{0.7935} & 0.6514 & \textbf{0.7957}\\
    & F-score & 0.6415 & \textbf{0.7134} & 0.6794 & \textbf{0.7116} & 0.6405 & \textbf{0.7104} & 0.6623 & \textbf{0.7194}\\
    & Jaccard & 0.4725 & \textbf{0.5544} & \textbf{0.5534} & 0.5527 & 0.5173 & \textbf{0.5502} & 0.4954 & \textbf{0.5625}\\
    \bottomrule
    \end{tabular}
    }
\end{table*}

As mentioned in Section \ref{introduction}, the communities identified by Bespoke, SEAL, CLARE, and ProCom may not contain some given nodes. Therefore, we can only compare PPSL with these methods alone, using only the communities found by these methods that contain the given nodes. Table \ref{table: global results} shows the experimental results of Bespoke, SEAL, CLARE, ProCom, and PPSL on five datasets.

The results in Table \ref{table: global results} demonstrate that PPSL outperforms Bespoke, SEAL, CLARE, and ProCom on most datasets. In particular, PPSL achieves superior performance over SEAL on Amazon, DBLP, Twitter, and YouTube datasets, and its F-score and Jaccard values consistently surpass those of Bespoke, CLARE, and ProCom across all datasets. 
The relatively poor performance of Bespoke, SEAL, CLARE, and ProCom stems from their global search paradigms, which learn general community structures but fail to capture the structural characteristics of communities specific to the query node.
As a result, the identified communities may not accurately reflect the ground-truth community structure of the query node. Additionally, SEAL achieves good performance on the LiveJournal dataset, primarily due to its generative adversarial network-based discriminator, which maintains high-quality community partitioning in the presence of large-scale and structurally diverse communities.

\subsection{Time of execution}

\begin{figure*}[!t]

  \centering
    \captionsetup[subfigure]{skip=0pt}
    \hspace{10mm}  
    \begin{subfigure}[b]{0.5\textwidth} 
        \centering
        \captionsetup{belowskip=-4pt, aboveskip=-4pt}  
        \includegraphics[width=\linewidth]{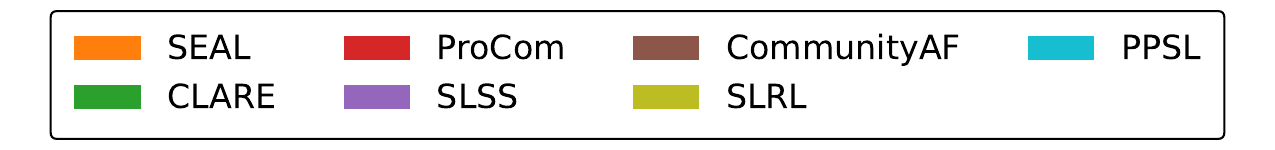}
        \caption{}
        \label{fig: methods}
    \end{subfigure}

    \begin{subfigure}[b]{0.31\textwidth}
        \includegraphics[width=\linewidth]{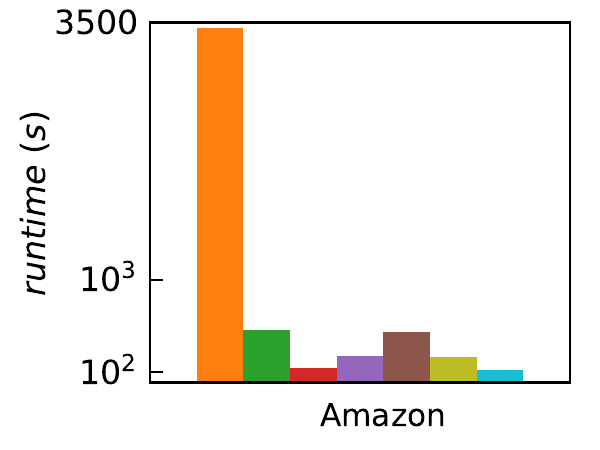}
        \caption{\hspace{1.2cm} (a)}
        \label{fig:e_sub1}
    \end{subfigure}
    \begin{subfigure}[b]{0.31\textwidth}
        \includegraphics[width=\linewidth]{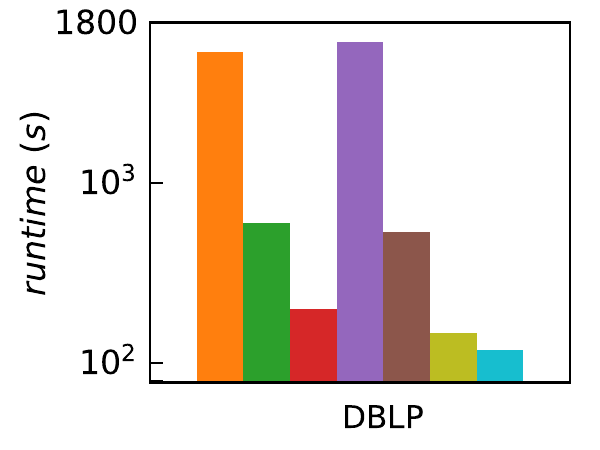}
        \caption{\hspace{1.3cm}(b)}
        \label{fig:e_sub2}
    \end{subfigure}
    \begin{subfigure}[b]{0.31\textwidth}
        \includegraphics[width=\linewidth]{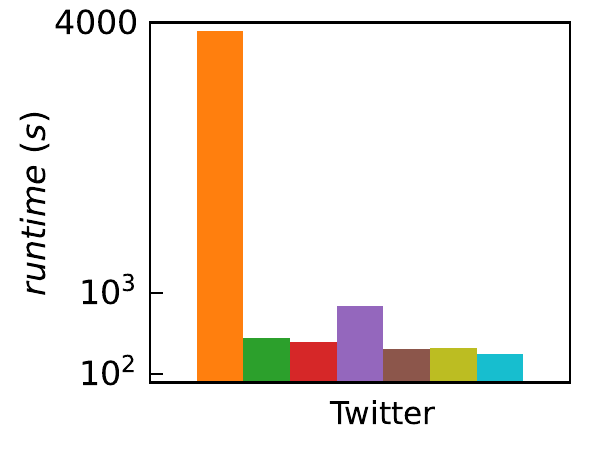}
        \caption{\hspace{1.3cm}(c)}
        \label{fig:e_sub3}
    \end{subfigure}
    
  \caption{Results of search efficiency.} 
  \label{figure: efficiency}
  
\end{figure*}

We compare the average runtime of PPSL with that of several semi-supervised community detection and semi-supervised local community detection methods on Amazon, DBLP, and Twitter, as shown in Figure \ref{figure: efficiency} (a), (b), and (c). Each experiment uses 100 query nodes, with the average time recorded as the algorithm’s runtime.

The results show that PPSL is the fastest among all semi-supervised local community detection methods (SLSS, CommunityAF, and SLRL), achieving nearly 37\% higher efficiency. The reasons are as follows: (1) The incremental GNN updates in CommunityAF are time-consuming. (2) PPSL avoids the costly matching of similar communities during community expansion, as required by SLSS, thus significantly reducing computational overhead. (3) Compared to SLRL, which performs more frequent community expansion during training, PPSL requires fewer expansion steps, thereby reducing the overall runtime.

In addition, PPSL is more runtime-efficient than semi-supervised community detection methods such as SEAL, CLARE, and ProCom. Specifically, PPSL differs from SEAL and CLARE in that its encoding process focuses on the local structural features of nodes. Moreover, during sample generation and prompt-driven fine-tuning, PPSL accesses only the neighbors of the query node, thereby reducing computational overhead. Although the runtime of ProCom is comparable to that of PPSL, the latter conducts local community detection based on the structure of the query node’s community, which incurs a higher computational cost but results in better community quality.

\subsection{Ablation study}

\begin{table*}[!t]
  \centering
  \caption{Results of ablation study. }
  \label{table: ablation}
  \small
  \setlength{\tabcolsep}{6mm}{
  \begin{tabular}{cccccc}
    \toprule
    Datasets & Metrics & w/o. NE & w/o. SG & w/o. PF & PPSL \\
    \midrule
    \multirow{4}{*}{\makecell{\textit{Amazon}}} & Precision & 0.7722 & 0.7813 & 0.7604 & \textbf{0.8683}\\
    \multirow{4}{*}{\textit{}} & Recall & 0.9137 & 0.9226 & 0.9174 & \textbf{0.9341} \\
    \multirow{4}{*}{\textit{}} & F-score & 0.8375 & 0.8465 & 0.8314 & \textbf{0.8827} \\
    \multirow{4}{*}{\textit{}} & Jaccard & 0.7197 & 0.7334 & 0.7112 & \textbf{0.8151} \\
    \midrule

    \multirow{4}{*}{\makecell{\textit{DBLP}}} & Precision & 0.4927 & 0.5109 & 0.4638 & \textbf{0.6833}\\
    \multirow{4}{*}{\textit{}} & Recall & 0.7033 & 0.7192 & 0.7143 & \textbf{0.7401} \\
    \multirow{4}{*}{\textit{}} & F-score & 0.5796 & 0.5977 & 0.5623 & \textbf{0.6684} \\
    \multirow{4}{*}{\textit{}} & Jaccard & 0.4075 & 0.4254 & 0.3913 & \textbf{0.5492} \\
    \midrule

    \multirow{4}{*}{\makecell{\textit{Twitter}}} & Precision & 0.3023 & 0.3114 & 0.3083 & \textbf{0.3395}\\
    \multirow{4}{*}{\textit{}} & Recall & 0.5486 & 0.5537 & 0.5544 & \textbf{0.5984} \\
    \multirow{4}{*}{\textit{}} & F-score & 0.3784 & 0.3855 & 0.3843 & \textbf{0.3943} \\
    \multirow{4}{*}{\textit{}} & Jaccard & 0.2424 & 0.2473 & 0.2502 & \textbf{0.2678} \\
    \midrule

    \multirow{4}{*}{\makecell{\textit{Youtube}}} & Precision & 0.4183 & 0.4224 & 0.4162 & \textbf{0.4495}\\
    \multirow{4}{*}{\textit{}} & Recall & 0.2696 & 0.2754 & 0.2623 & \textbf{0.3802} \\
    \multirow{4}{*}{\textit{}} & F-score & 0.2833 & 0.2724 & 0.2865 & \textbf{0.3177} \\
    \multirow{4}{*}{\textit{}} & Jaccard & 0.1694 & 0.1636 & 0.1723 & \textbf{0.2142} \\
    \midrule

    \multirow{4}{*}{\makecell{\textit{LiveJournal}}} & Precision & 0.4216 & 0.4477 & 0.4134 & \textbf{0.5842}\\
    \multirow{4}{*}{\textit{}} & Recall & 0.6793 & 0.6924 & 0.6894 & \textbf{0.8004} \\
    \multirow{4}{*}{\textit{}} & F-score & 0.5203 & 0.5435 & 0.5167 & \textbf{0.6174} \\
    \multirow{4}{*}{\textit{}} & Jaccard & 0.3513 & 0.3734 & 0.3481 & \textbf{0.4974} \\
    
  \bottomrule
\end{tabular}
}
\end{table*}

To evaluate the effectiveness of each component in PPSL, we conduct ablation studies on five datasets and design three simplified variants of the model: (1) “w/o. NE”: The node encoding component is removed, and node representations are directly derived from their original feature vectors; (2) “w/o. SG”: The sample generation component is removed, and the 2-ego subgraph of the given node is used as its initial community instead; (3) “w/o. PF”: The prompt-driven fine-tuning component is removed, and the initial community is directly treated as the final prediction community. We select 100 nodes as the given nodes to conduct experiments.

Table \ref{table: ablation} presents the performance of these three simplified variants and the complete PPSL model across the datasets. Overall, PPSL consistently outperforms w/o. NE, w/o. SG, and w/o. PF, confirming the importance of the node encoding, sample generation, and prompt-driven fine-tuning components. More specifically, PPSL surpasses w/o. NE, indicating that the node encoding phase helps the model capture the features of local structures, which benefits downstream tasks. Compared to w/o. SG, the superior performance of PPSL demonstrates that high-quality initial communities lead to better training community sets, thereby enhancing the quality of prompt-based learning. Finally, compared to w/o. PF, PPSL achieves better performance in the metrics of F-score and Jaccard, demonstrating that using a training community set tailored to the given node as prompts can effectively enhance the quality of community prediction. Moreover, the prompt-based learning mechanism enables the model to capture the structural characteristics of the given node's community.

\subsection{Parameter study}

\begin{figure}[!t]
\centering
    \captionsetup[subfigure]{skip=0pt}
    \hspace{7mm}  
    \begin{subfigure}[b]{0.65\columnwidth} 
        \centering
        \captionsetup{belowskip=-3pt, aboveskip=-3pt}  
        \includegraphics[width=\linewidth]{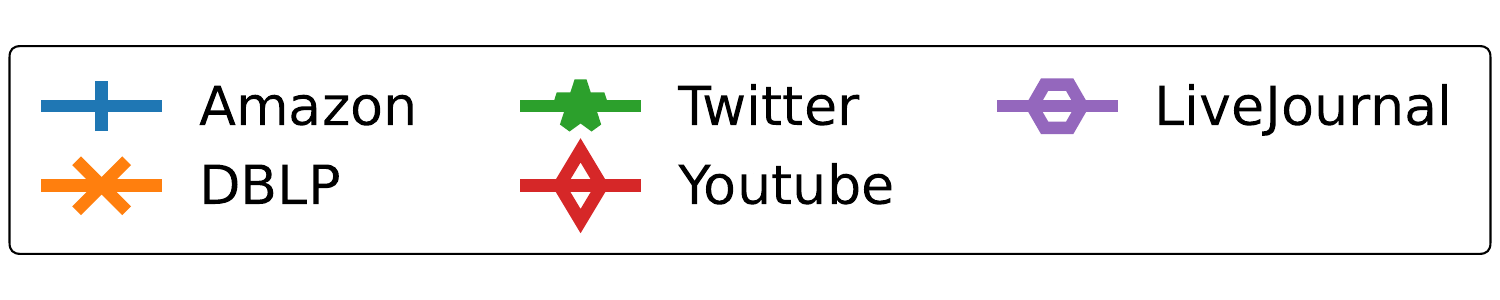}
        \caption{}
        \label{fig: datasets}
    \end{subfigure}

    \begin{subfigure}[b]{0.7\columnwidth}
        \includegraphics[width=\linewidth]{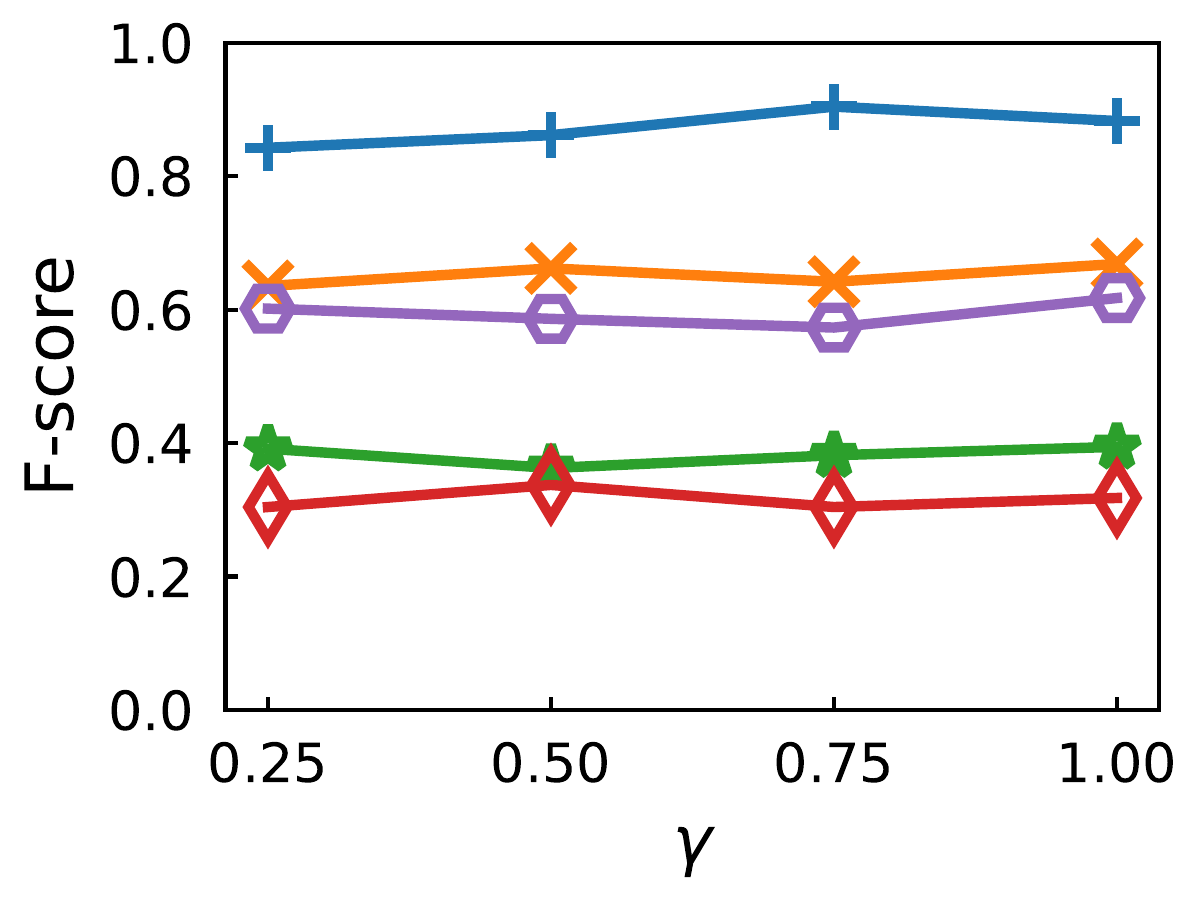}
        \caption{\hspace{0.9cm}(a) Effect of the discount factor $\gamma$}
        \label{fig: p_sub1}
    \end{subfigure}
    \hfill  
    \begin{subfigure}[b]{0.7\columnwidth}
        \includegraphics[width=\linewidth]{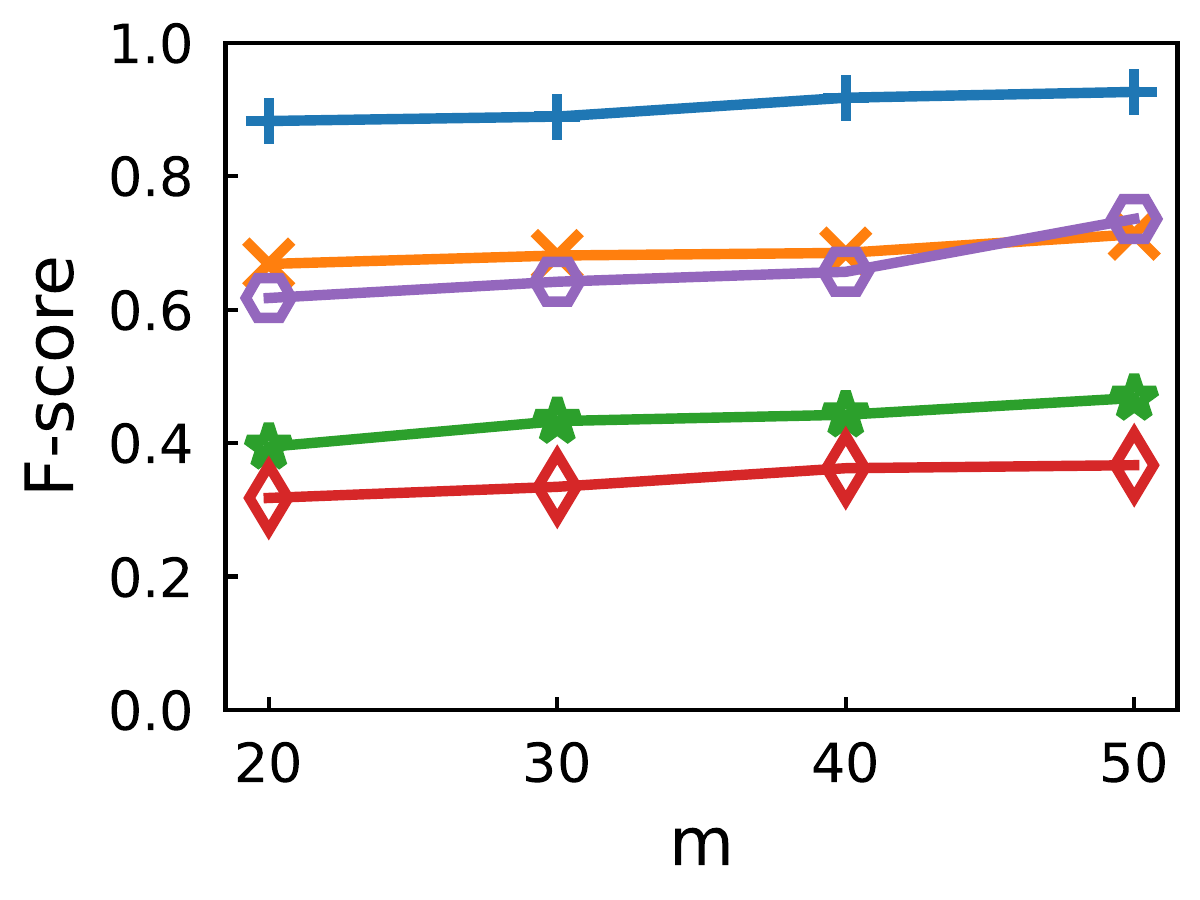}
        \caption{\hspace{1.0cm}(b) Effect of sample numbers $m$}
        \label{fig: p_sub2}
    \end{subfigure}
    \caption{Results of PPSL parameter experiments. }
    \label{figure: parameter}
\end{figure}

We analyze the effect of the hyperparameters $\gamma$ and $m$ on the performance of PPSL. $\gamma$ is the discount factor for computing cumulative rewards during sample generation, chosen from \{0.25, 0.5, 0.75, 1\}. $m$ is the number of samples used for prompt-driven fine-tuning, with values in \{20, 30, 40, 50\}. Specifically, we vary one parameter while keeping the others fixed and evaluate the changes in F-score across five datasets. The results are shown in Figure \ref{figure: parameter} (a) and (b).

(1) Effect of $\gamma$: On Amazon, the performance of PPSL steadily improves as $\gamma$ increases from 0.25 to 0.75, then slightly declines when $\gamma$ reaches 1. 
For DBLP, Twitter, and YouTube, PPSL's performance fluctuates slightly when $\gamma$ increases from 0.25 to 0.75, but peaks at $\gamma = \text{1}$.
For LiveJournal, performance decreases as $\gamma$ increases from 0.25 to 0.75, and then slightly improves at $\gamma = \text{1}$.
We set $\gamma=\text{1}$ for other experiments.

(2) Effect of $m$: Even when only 20 similar samples are provided for each query node, PPSL achieves strong performance across all datasets. While performance generally improves with increasing $m$, the overall change from $m = \text{20}$ to $m = \text{50}$ remains modest. The default value of $m$ during the experiments is 20.

\section{Conclusion}  \label{conclusion}
This paper proposes Pre-trained Prompt-driven Semi-supervised Local community detection (PPSL), the first attempt to apply the “pre-train, prompt” paradigm to semi-supervised local community detection. Unlike existing methods designed for semi-supervised community detection, PPSL focuses on identifying the community of a given node with improved accuracy and efficiency. The model integrates three core components: a node encoding component that learns node and community representations through local structural patterns using graph neural networks; a sample generation component that constructs training data by identifying an initial community for a given node and retrieving structurally similar communities; and a prompt-driven fine-tuning component that uses training samples as prompts to guide the final community prediction. Extensive experiments on five real-world datasets show that PPSL consistently outperforms baseline methods. Ablation studies validate the contribution of each component. In future work, we aim to further integrate the “pre-train, prompt” paradigm with large language models to enhance its applicability to more complex network analysis tasks.

\section*{Acknowledgments}
This work was supported by the National Natural Science Foundation of China [No.62106004, No.62272001, and No.62206004]. 

\bibliographystyle{IEEEtran}
\bibliography{references}

\vfill

\end{document}